\documentclass[12pt,a4paper]{article}
\usepackage[utf8x]{inputenc}
\usepackage{amsmath}
\usepackage{graphicx}
\usepackage[left=2cm,right=2cm,top=2cm,bottom=2cm]{geometry}
\usepackage[bookmarks=true,colorlinks=true,citecolor=blue,linkcolor=red,urlcolor=magenta]{hyperref}

\usepackage[numbers,sort&compress]{natbib}

\title{Impulse source of high energy neutrons emitted by fusion reactions after
       compression of D-T gas by cumulative detonation waves}%

\author{V.D.~Rusov\footnote{Corresponding author e-mail: siiis@te.net.ua},
        V.A.~Tarasov, S.A.~Chernezhenko,\\ V.P.~Smolyar, V.V.~Urbanevich,
        T.N.~Zelentsova}

\date{}

\begin{document}

\maketitle

\begin{center}
    \textit{Odessa National Polytechnic University,\\
    Department of Theoretical and Experimental Nuclear Physics,\\ 
    Shevchenko av.~1, 65044 Odessa, Ukraine}
\end{center}

\begin{abstract}
We develop the physical model and the system of equations for the impulse
neutron source (INS) of high-energy neutrons ($\sim$ 14 MeV) emitted by fusion
reactions during compression of D-T gas by cumulative detonation waves. The
system of INS equations includes a system of gas dynamic equations that takes
into account the energy transfer by radiation, equations for the radiation
flux, the equation of the shock adiabat (the Hugoniot adiabat) for a compressed
gas, and the equation for the neutron yield.

We perform the INS dynamics simulation for the spherical and cylindrical
geometries, and calculate maximum temperatures of D-T plasma, its density
and neutron yield in the pulse.

The obtained temperature estimates and simulation results show that the
thermonuclear fusion temperatures are reached within this approach, and the
fusion reactions proceed. Their yield determines the yield of neutrons.
\end{abstract}

\section{Introduction}

In a recent book by Kozyrev~\cite{1} it was proposed to conduct the research
and development work ($R \& D$), to develop a gas dynamic thermonuclear neutron
source (GDTS). The Author suggested to initiate a thermonuclear fusion reaction
$D\left( {d,n} \right){}^3He$
by dynamically compressing a deuterium-deuterium (D-D) gas mixture with a
spherical converging detonation wave (cumulative detonation wave), which would
give rise to a neutron flux. Some estimates were also made, which justify the
principal feasibility of such source of neutrons.

%In \cite{1}, it was proposed to initiate a thermonuclear fusion reaction $D\left( {d,n} \right){}^3He$ by dynamically compressing a deuterium-deuterium (D-D) gas mixture by a spherical converging detonation explosion wave (cumulative detonation wave), which would create a neutron flux. Some estimates are also presented in [1], which justify the principal possibility of creating such source of neutrons.

Such $R \& D$ work was apparently carried out, since the preface to~\cite{1} as
well as~\cite{2,3,4,5} mention the experiment conducted in 1982 in VNIIEF
(Sarov, Russia) involving a device built on the basis of Kozyrev approach,
using a liquid explosive (tetranitromethane in nitrobenzene). The neutron yield
was reported to reach $4 \cdot {10^{13}}$ neutrons per pulse. According
to~\cite{4}, in this device a deuterium-tritium (D-T) gas with the initial
radius of ${r_0}= 1 mm$ and initial density of ${\rho_0}= 0.1 ~g/cm^3$
underwent a compression. The temperature reached T = 0.65~keV
($T = 0.78 \cdot 10^7~K$), and the maximum density
${\rho _{\max }} = 80~g/cm^3$.

Still, according to~\cite{4,5}, the yield of neutrons turned out to be 2-3
orders of magnitude lower than expected ($9.5 \cdot {10^{16}}$). As noted
in~\cite{4,5}, this deviation may be associated with asymmetry and mixing at
the boundary of substances of different density, and increasing gas dynamic
instabilities in GDTS.

Let us note that the temperature of a thermonuclear plasma was apparently
estimated by a method described in~\cite{6}: the neutron spectrum was measured
during the experiment, and the temperature was calculated from the half-width
of the neutron spectrum maximum.

In~\cite{7,7a} a phenomenon of the unlimited cumulation not related to a
centripetal motion of the gas (i.e. for a plane shock wave) was theoretically
discovered. It was caused by a special periodic structure of matter ("sandwich")
along the shock wave path. Such motion is a periodic self-similar process.

Here we present the estimates that also confirm the possibility
of the impulse neutron source (INS) implementation with the neutrons being
emitted by fusion reactions $T\left( {d,n} \right){}^4He$ during the
compression of the D-T gas by cumulative detonation waves. We also define the
physical model and a system of equations (gas dynamic and other necessary
equations for the model) for the INS. On the basis of this model we developed a
computer software and conducted the numerical simulations. Our simulation
results demonstrate a good agreement with preliminary estimates.

\section{Estimation of the INS feasibility using D-T gas compression by cumulative detonation waves}

In order to substantiate the fundamental possibility of INS realization by
the thermonuclear fusion reactions $T\left( {d,n} \right){}^4He$ under
compression of D-T gas by cumulative detonation waves, the following
preliminary estimates may be given.

The attainable temperatures for the compression of a deuterium-tritium gas by
spherical convergent shock waves can be calculated within the framework of the
Zel'dovich model~\cite{8}. It deals with the adiabatic compression of a gaseous
medium by a running plane shock wave, and gives the expressions for the maximum
compression of a gas and for the dependence of its temperature on the amplitude
of a shock wave pressure:

\begin{equation}
\frac{{{V_0}}}{{{V_1}}} = \frac{{\gamma  + 1}}{{\gamma  - 1}}, 
\label{eq1}
\end{equation}

\noindent
where ${V_0}$ is the initial volume, ${V_1}$ is the volume of the gas after
compression, $\gamma $ is the adiabatic index,

\begin{equation}
\frac{{{T_1}}}{{{T_0}}} = \frac{{\gamma  - 1}}{{\gamma  + 1}} \cdot
\frac{{{P_1}}}{{{P_0}}}, 
\label{eq2}
\end{equation}

\noindent
where ${T_0},{P_0}$ are the initial temperature and pressure of the gas,
${T_1}$ is the maximum temperature of the gas after its compression by a shock
wave, and ${P_1}$ is the amplitude of the shock wave pressure.

It was also shown in~\cite{8} that the gas compression is described by the
Hugoniot state equation (shock adiabat), which depends on two parameters (e.g.
$f(T,V,P)$, where $V$ and $P$ are parameters) in contrast to the Clapeyron
equation, which depends on a single parameter (e.g. $f(T,V)$, where $V$ is a
parameter).

So for the maximum temperature of the gas being compressed by a spherical
convergent shock wave, we obtain the following expression:

\begin{equation}
{T_1} = {c_{coll}} \cdot {c_{sph}} \cdot \frac{{\gamma  - 1}}{{\gamma  + 1}} \cdot \frac{{{P_1}}}{{{P_0}}} \cdot {T_0},	
\label{eq3}
\end{equation}

\noindent
where ${c_{sph}}$ is the coefficient that takes into account the growth of the
pressure amplitude in a converging spherical shock wave and which, using
(\ref{eq1}), may be expressed as:

\begin{equation}
{c_{sph}} = {\left( {\frac{{{r_0}}}{{{r_1}}}} \right)^2} = {\left( {\frac{{{V_0}}}{{{V_1}}}} \right)^{\frac{2}{3}}} = {\left( {\frac{{\gamma  + 1}}{{\gamma  - 1}}} \right)^{\frac{2}{3}}},
\label{eq4}
\end{equation}

\noindent
where ${r_0}$ is the initial radius of the gas having a spherical shape,
${r_1}$ is the minimum radius of the compressed gas,
and ${c_{coll}}$ is the coefficient that takes into account the increase in the
pressure amplitude in a convergent spherical shock wave caused by the collision
of a converging spherical wave\footnote{An analogue of the increase in kinetic
energy transformed into the energy of reaction products after the particles
collision in the center-of-mass system (C-system) as compared to the reaction
with a stationary target, i.e. in a laboratory system (L-system)}, which,
according to~\cite{1}, can be chosen equal to 10.

It is well known that the maximum energy yield in an exothermic nuclear
reaction is in the C-system. However, few researchers pay attention to the
actual gain in the reaction energy when carried out in the C-system (e.g. in
the collider) relative to the L-system (stationary target exposed to the
particle beam). A good illustration of such gain is given in~\cite{9}, which
gives the minimum kinetic energy of colliding protons and antiprotons,
necessary for the formation of the neutral bosons of a weak field ($Z^0$
bosons) in C-system and L-system. The minimum kinetic energy for the protons
is 45.6~GeV for the C-system, and 4434.0~GeV for the L-system. So the gain in
the energy of the reaction products is 100 times.

According to~\cite[p.56]{1} and~\cite{10}, when two shock waves with equal
pressure amplitude collide ($N = 2$), ${c_{coll}} = 6$, and in the case of a
converging spherical wave $N > 2$ and ${c_{coll}} = 10$.

According to~\cite{8}, 
%$\gamma  = {5 \mathord{\left/  {\vphantom {5 3}} \right. \kern-\nulldelimiterspace} 3}$
$\gamma = 5 / 3$ and the maximum compression is 4 for a monatomic gas,
%$\gamma  = {7 \mathord{\left/  {\vphantom {7 5}} \right. \kern-\nulldelimiterspace} 5}$
$\gamma = 7/5$  and the maximum compression is 6 for a diatomic gas with no
excitations, and 
%$\gamma  = {9 \mathord{\left/  {\vphantom {9 7}} \right. \kern-\nulldelimiterspace} 7}$
$\gamma = 9/7$ and the maximum compression is 8 for a diatomic gas with
 excitations.

\begin{figure}[tb!]% figure* for wide figure, [h] [!] to change the placement
\centering
\includegraphics[width=8cm]{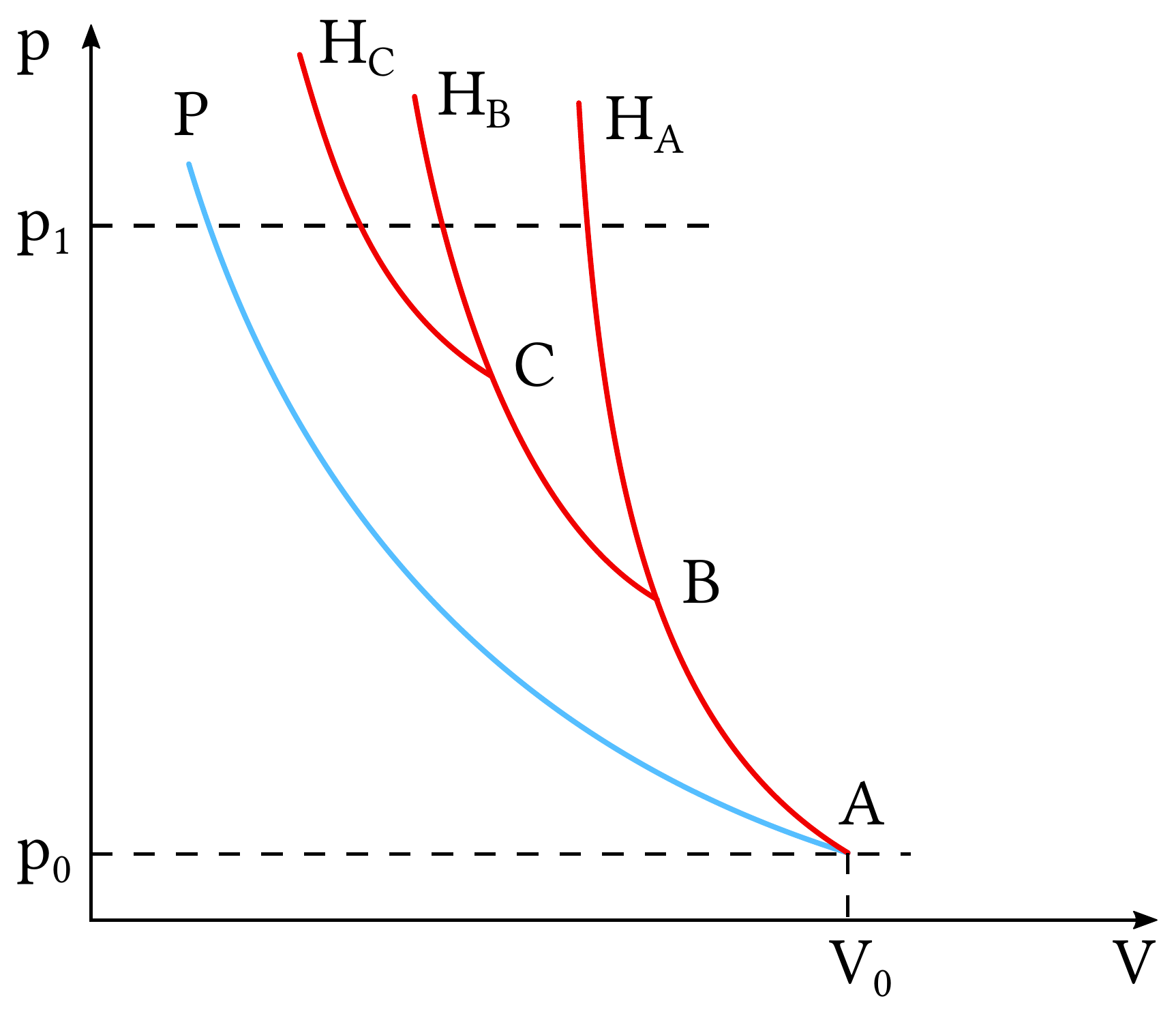}
\vskip-3mm
\caption{On the issue the single and multiple adiabatic compressions of a gas
to the same pressure $P_1$ ( $H_A$, $H_B$, $H_C$ are the shock adiabats for
which $A$, $B$ and $C$ are the initial points; $P$ is the Poisson
adiabat)~\cite{8}.}
\label{fig1}
\end{figure}

Since the Hugoniot equation depends on two parameters, it is impossible to
reach the same finite state of the gas by compressing with several shock waves
and a single shock wave, starting from the same initial state~\cite{8}. For
example, if a strong shock wave passes through a monoatomic gas, the gas will be
compressed by four times. By contrast, if two consecutive strong shock waves
(with the same final pressure) pass through the gas, it will be compressed by
16 times. This is demonstrated well in Fig.~\ref{fig1} from~\cite{8}.

Let us remark that such compression of a gas by several consecutive spherical
convergent shock waves is called a "cascade shock wave" by Kozyrev in~\cite{1}.

The data on the amplitudes of the pressure in shock waves are given in
Table~\ref{tab1} for several classical powerful explosives (pentaerythritol
tetranitrate (PETN), nitroglycerine, lead azide), and may be used for the
estimations.

\begin{table*}[tbp!]
\noindent\caption{Data from~\cite{1}}\vskip3mm%\tabcolsep4.2pt
\label{tab1}
\footnotesize
\centering
\begin{tabular}{|c|c|c|c|c|c|}
 \hline%
 \rule{0pt}{5mm}
   \parbox[t]{2cm}{Explosive}%
 & \parbox[t]{1.7cm}{Detonation velocity, m/s}
 & \parbox[t]{2cm}{Explosive density, $g/sm^3$}
 & \parbox[t]{2cm}{Detonation products density, $g/sm^3$ }
 & \parbox[t]{2.5cm}{Detonation products pressure, $P_D$, atm}
 & \parbox[t]{2cm}{Pressure at reflection, $P_R$, atm}\\[2mm]%
\hline%
PETN & 7900 & 1.60 & 2.12 & 250000 & 560000 \\
\hline
Nitroglycerine & 7900 & 1.60 & 2.12 & 250000 & 560000\\%
\hline
Lead azide & 5890 & 4.70 & 6.30 & 400000 & 900000\\
\hline
\end{tabular}
\end{table*}

So using (\ref{eq3}) for ${T_0} = 300~K, ~{P_0} = 1~atm.$, we estimate the
maximum temperatures. They are presented in Tables~\ref{tab2} and~\ref{tab3}.

\begin{table*}[tbp!]
\noindent\caption{Calculated data for $c_{coll}=1$}\vskip3mm\tabcolsep4.2pt
\label{tab2}
\footnotesize
\centering
\begin{tabular}{|c|c|c|c|c|c|c|c|}
 \hline%
 \rule{0pt}{5mm}
   \parbox[t]{1.5cm}{Explosive}%
 & \parbox[t]{1.5cm}{Pressure $P_1,$ atm}
 & \parbox[t]{1.8cm}{Single shock wave}
 & \parbox[t]{1.8cm}{Cascade of two shock waves }
 & \parbox[t]{1.7cm}{Monatomic gas, $c_{sph}$}
 & \parbox[t]{2.0cm}{Diatomic gas without excitations, $c_{sph}$}
 & \parbox[t]{2.0cm}{Diatomic gas with excitations, $c_{sph}$}
 & \parbox[t]{1.8cm}{Maximum temperature, $10^7~K$}\\[2mm]%
\hline%
PETN       & 250000 & $\bullet$ & &2,25&-&-&4,22\\
\hline
PETN       & 250000 & & $\bullet$ &6,34&-&-&11,89\\%
\hline
PETN       & 250000 & $\bullet$ & &-&3,30&-&4,21\\%
\hline
PETN       & 250000 & & $\bullet$ &-&10,88&-&13,87\\%
\hline
PETN       & 250000 & $\bullet$ & &-&-&3,99&3,89\\%
\hline
PETN       & 250000 & & $\bullet$ &-&-&15,96&15,56\\%
\hline
Lead azide & 400000 & $\bullet$ & &2,25&-&-&6,75\\%
\hline
Lead azide & 400000 & & $\bullet$ &6,34&-&-&19,02\\%
\hline
Lead azide & 400000 & $\bullet$ & &-&3,30&-&6,74\\%
\hline
Lead azide & 400000 & & $\bullet$ &-&10,88&-&22,19\\%
\hline
Lead azide & 400000 & $\bullet$ & &-&-&3,99&6,24\\%a
\hline
Lead azide & 400000 & & $\bullet$ &-&-&15,96&24,90\\%
\hline
\end{tabular}
\end{table*}

\begin{table*}[tbp!]
\noindent\caption{Calculated data for $c_{\text{coll}}=2$}\vskip3mm\tabcolsep4.2pt
\label{tab3}
\footnotesize
\centering
\begin{tabular}{|c|c|c|c|c|c|c|c|}
 \hline%
 \rule{0pt}{5mm}
   \parbox[t]{1.5cm}{Explosive}%
 & \parbox[t]{1.5cm}{Pressure $P_1,$ atm}
 & \parbox[t]{1.8cm}{Single shock wave}
 & \parbox[t]{1.8cm}{Cascade of two shock waves }
 & \parbox[t]{1.7cm}{Monatomic gas, $c_{\text{sph}}$}
 & \parbox[t]{2.0cm}{Diatomic gas without excitations, $c_{\text{sph}}$}
 & \parbox[t]{2.0cm}{Diatomic gas with excitations, $c_{\text{sph}}$}
 & \parbox[t]{1.8cm}{Maximum temperature, $10^7~K$}\\[2mm]%
\hline%
PETN       & 250000 & $\bullet$ & &2,25&-&-&8,44\\ % �������/Example: 1&2&3&4&5\\
\hline
PETN       & 250000 & & $\bullet$ &6,34&-&-&23,78\\%
\hline
PETN       & 250000 & $\bullet$ & &-&3,30&-&8,42\\%
\hline
PETN       & 250000 & & $\bullet$ &-&10,88&-&27,74\\%
\hline
PETN       & 250000 & $\bullet$ & &-&-&3,99&7,78\\%
\hline
PETN       & 250000 & & $\bullet$ &-&-&15,96&31,12\\%
\hline
Lead azide & 400000 & $\bullet$ & &2,25&-&-&13,50\\%
\hline
Lead azide & 400000 & & $\bullet$ &6,34&-&-&38,04\\%
\hline
Lead azide & 400000 & $\bullet$ & &-&3,30&-&13,48\\%
\hline
Lead azide & 400000 & & $\bullet$ &-&10,88&-&44,38\\%
\hline
Lead azide & 400000 & $\bullet$ & &-&-&3,99&12,48\\%
\hline
Lead azide & 400000 & & $\bullet$ &-&-&15,96&49,80\\%
\hline
\end{tabular}
\end{table*}

These estimates show that the thermonuclear temperatures may be reached.

%\section{PHYSICAL MODEL OF INS, SYSTEM OF GAS DYNAMIC EQUATIONS AND OTHER EQUATIONS OF INS AND RESULTS OF MODELING}
\section{Physical model of the INS and the simulation results}

The physical model of the
%IIN -- WHAT?
INS should include the heat transfer equation in addition to the hydrodynamic
equations (gas dynamics).

Because of the high-speed processes in the INS (according to Table~\ref{tab1},
the shock wave velocity is $v \sim$ 6000-8000 m/s), it is necessary to take
into account the radiation energy losses. Indeed, a qualitative switch to the
radiation energy transfer happens when the hydrodynamic and radiation energy
fluxes become comparable~\cite{11}, i.e. under the condition:

\begin{equation}
%{{\sigma {T^4}} \mathord{\left/ {\vphantom {{\sigma {T^4}} {\rho {V^3}}}} \right. \kern-\nulldelimiterspace}{\rho {V^3}}} \sim 1,
\sigma {T^4} / \rho {v^3} \sim 1,
\label{eq5}
\end{equation}

\noindent
where $\sigma$ is the Stefan-Boltzmann constant, $\rho $ is the medium
density.

The system of gas dynamics equations for INS in the Euler representation
(e.g~\cite{8,11}) is:

\begin{equation}
\frac{{\partial \rho }}{{\partial t}} + div\left( {\rho \vec V\left( {\vec r,t} \right)} \right) = 0,
\label{eq6}
\end{equation}

\begin{equation}
\frac{{\partial \vec V\left( {\vec r,t} \right)}}{{\partial t}} + (\vec V\left( {\vec r,t} \right)div)\vec V\left( {\vec r,t} \right) + \frac{1}{\rho }\nabla P\left( {\vec r,t} \right) = 0,
\label{eq7}
\end{equation}

\begin{equation}
\frac{{\partial \rho E}}{{\partial t}} + div\left( {\rho E\vec V} \right) + div\left( {P\vec V} \right) + \rho Q - div\vec S = 0,
\label{eq8}
\end{equation}

\noindent
where $\rho $ is the medium density, $\vec V$ is the mass velocity, 
%$E = \varepsilon  + {{{V^2}} \mathord{\left/ {\vphantom {{{V^2}} 2}} \right. \kern-\nulldelimiterspace} 2}$
$E = \varepsilon  + V^2 / 2$ is the the total energy density, and $\varepsilon$
is its thermal component, $P$ is the pressure, $Q$ is the energy release
per unit time (proportional to the fusion reaction rate in the compressed D-T
plasma), $\vec S$ is the radiation flux.

Equation for the radiation flux:

\begin{equation}
\vec S\left( {\vec r,t} \right) = \int\limits_0^\infty  {{{\vec S}_\nu }} d\nu ,
\label{eq9}
\end{equation}

\noindent where

\begin{equation}
{\vec S_\nu } = \int\limits_0^\infty  {{I_\nu }} \vec \Omega d\Omega ,
\label{eq10}
\end{equation}

\begin{equation}
{I_\nu }\left( {\vec r,\vec \Omega ,t} \right) = \hbar \nu cf\left( {\nu ,\vec r,\vec \Omega ,t} \right)d\nu d\Omega ,
\label{eq11}
\end{equation}

\noindent
where $\vec \Omega$ is the unit vector specifying the direction of the quanta
propagation, $\nu$ is the quanta frequency, $c$ is the speed of light in vacuum,
$\hbar $ is the Planck constant, $f$ is the frequency distribution function of
the radiation intensity.

In order to specify the frequency distribution of the radiation intensity, let
us use the Wien function, e.g.~\cite{12}:

\begin{equation}
%{f_\nu } = \frac{{8\pi {\nu ^3}}}{{{c^3}}}\exp \left( { - {{\hbar \,2\pi \,\nu } \mathord{\left/ {\vphantom {{\hbar \,2\pi \,\nu } {kT}}} \right. \kern-\nulldelimiterspace} {kT}}} \right).	
f_\nu = \frac{8\pi \nu ^3}{c^3} \exp \left( -\hbar 2\pi \nu / kT \right).	
\label{eq12}
\end{equation}

To complete the problem, it is necessary to introduce the equation of state for
the compressed D-T gas. We use the shock adiabat (Hugoniot equation) for this
purpose~\cite{8,13,13a}:

\begin{equation}
{P_1} = H\left( {{V_1},{P_0},{V_0}} \right),
\label{eq13}
\end{equation}

\noindent
which may be expressed explicitly when the thermodynamic relations
$\varepsilon \left( {P,V} \right)$ or $\varepsilon \left( {P,\rho } \right)$
are expressed in simple equations. Let us note that the thermodynamic
properties of the system are assumed to be known in our problem.

For example, according to~\cite{8}, the Hugoniot equation for an ideal gas with
a constant heat capacity has the following form:

\begin{equation}
{P_1} = \frac{{\left( {\gamma  + 1} \right){V_0} - \left( {\gamma  - 1} \right){V_1}}}{{\left( {\gamma  + 1} \right){V_1} - \left( {\gamma  - 1} \right){V_0}}}{P_0},
\label{eq14}
\end{equation}

\noindent and the ratio of the volumes and temperatures:

\begin{equation}
\frac{{{V_1}}}{{{V_0}}} = \frac{{\left( {\gamma  - 1} \right){P_1} + \left( {\gamma  + 1} \right){P_0}}}{{\left( {\gamma  + 1} \right){P_1} + \left( {\gamma  - 1} \right){P_0}}},
\label{eq15}
\end{equation}

\begin{equation}
\frac{{{T_1}}}{{{T_0}}} = \frac{{{P_1}{V_1}}}{{{P_0}{V_0}}}.	
\label{eq16}
\end{equation}

We solved the system of gas dynamics equations for INS
((\ref{eq6})-(\ref{eq8})) with respect to unknown $P$, $\rho$ and $\vec V$
numerically. For this purpose we applied the particle-in-cell method~\cite{14}.
The radiation transfer equations in the multigroup approximation were solved
using the ${S_n}$-method~\cite{15}.

We tested this scheme by comparing the simulated deuterium compression to the
experimental $P - \rho $ diagram for deuterium~\cite{13,13a}.

The neutron yield is equal to the yield of the thermonuclear fusion reaction
$T\left( {d,n} \right){}^4He$, and was calculated as follows:

\begin{equation}
\varphi  = \int\limits_{{t_s}}^{{t_f}} {\tilde \rho \left\langle {v\sigma } \right\rangle } dt, 
\label{eq17}
\end{equation}

\noindent
where $\tilde \rho $ is the density of the plasma particles,
$\left\langle {v\sigma } \right\rangle $ is the fusion reaction cross section
averaged over the velocities of the thermal motion of the particles. According
to~\cite{16}, it was given as:

\begin{equation}
\left\langle {v\sigma } \right\rangle  = Const \cdot \theta ^{-\frac{2}{3}}
%{ - {2 \mathord{\left/ {\vphantom {2 3}} \right. \kern-\nulldelimiterspace} 3}}}
\exp \left[ - 3 \left(\frac{\pi {e^2}}{\hbar c}
% \mathord{\left/ {\vphantom {{\pi {e^2}} {\hbar c}}} \right. \kern-\nulldelimiterspace} {\hbar c}}}
  \right) ^{\frac{2}{3}}
% \mathord{\left/  {\vphantom {2 3}} \right. \kern-\nulldelimiterspace} 3}}}
 \left( \frac{m{c^2}}{2 \theta} 
% \mathord{\left/ {\vphantom {{m{c^2}} {2\theta }}} \right. \kern-\nulldelimiterspace} {2\theta }}}
 \right) ^{\frac{1}{3}}
% \mathord{\left/ {\vphantom {1 3}} \right. \kern-\nulldelimiterspace} 3}}}} 
\right],
\label{eq18}
\end{equation}

\noindent
where 
%$Const = {2^{{5 \mathord{\left/ {\vphantom {5 6}} \right. \kern-\nulldelimiterspace} 6}}}{3^{ - {1 \mathord{\left/  {\vphantom {1 2}} \right. \kern-\nulldelimiterspace} 2}}}\frac{{2\pi {e^2}}}{\hbar }{m^{{{ - 1} \mathord{\left/  {\vphantom {{ - 1} 3}} \right. \kern-\nulldelimiterspace} 3}}}$,
$Const = 2^{5/6} \cdot 3^{-1/2} \cdot \frac{2\pi {e^2}}{\hbar} \cdot {m^{-1/3}}$,
$\theta  = kT$, $m$ is the reduced mass of the interacting nuclides D and T,
$e$ is the elementary charge, ${t_s}$ is the time of the fusion reactions start
(the moment when the thermonuclear temperatures are reached in the compressed
D-T plasma), ${t_f}$ is the fusion reactions end time (determined as the moment
when the mass velocity becomes zero).

\begin{figure}[tb!]% figure* for wide figure, [h] [!] to change the placement
\centering
\includegraphics[width=8cm]{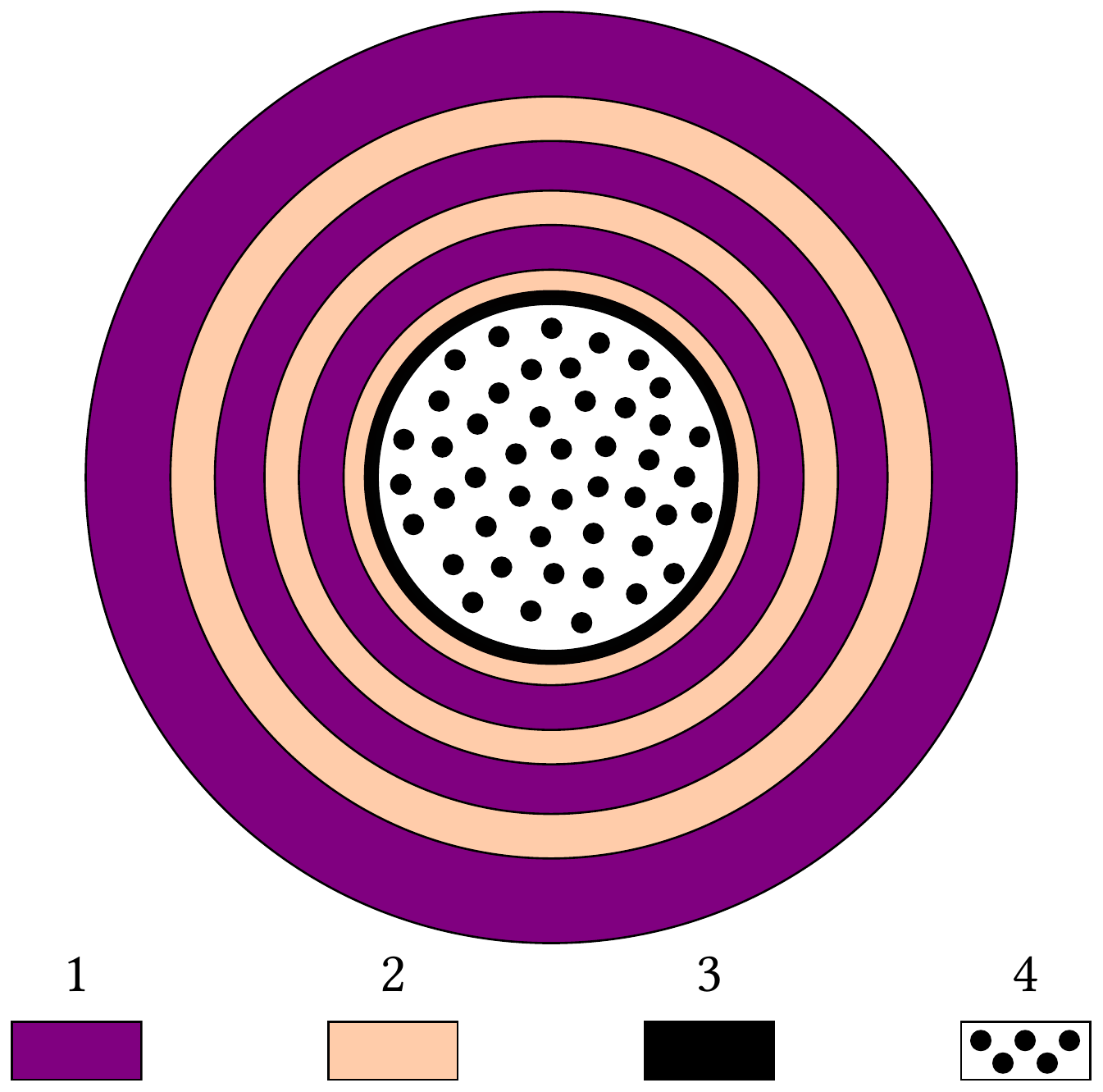}
\vskip-3mm\caption{A model of the spherical INS. (1 - explosive layers,
2 - Polymethylpentene layers, 3 - metal liner, 4 - gas mixture of D-T).}
\label{fig2}
\end{figure}

\begin{figure}[tb!]% figure* for wide figure, [h] [!] to change the placement
\centering
\includegraphics[width=8cm]{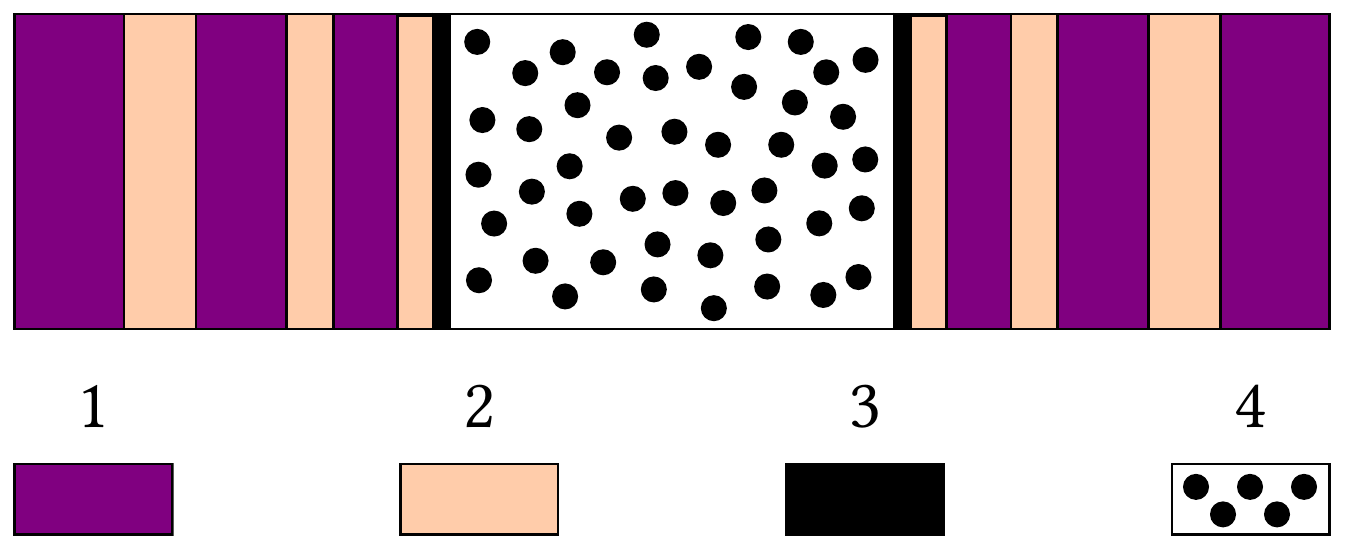}
\vskip-3mm\caption{A model of the cylindrical INS (1 - explosive layers,
2 - Polymethylpentene layers, 3 - metal liner, 4 - gas mixture of D-T).}
\label{fig3}
\end{figure}

We simulated spherical (Fig.~\ref{fig2}) and cylindrical (Fig. \ref{fig3}) INS. 

We performed the simulation of the INS with a single explosive layer, as well
as a "sandwich" structure of interlaced explosive (lead azide, 
$\rho = 4.7~g/cm^3$) and Polymethylpentene ($\rho = 0.8~g/cm^3$) layers.
One mole of D-T gas ($\rho  = 0.1~g/cm^3$) was compressed, and the mass of
the explosive was $\sim$500~kg. The spherical INS simulation demonstrated
the maximum temperature of $2 \cdot {10^8}~K$, the maximum density 
${\rho _{\max }}= 1.6~g/cm^3$, and the neutron yield of $9.4 \cdot {10^{17}}$
per pulse (neutron energy $\sim$ 14~MeV). The cylindrical INS demonstrated the
maximum temperature of $1.3 \cdot {10^8}~K$, the maximum density  
${\rho _{\max }}= 1.2~g/cm^3$ and the neutron yield of $6.7 \cdot {10^{16}}$
per pulse.

\newpage

\section{Conclusions}

We developed a physical model and a system of equations for the impulse
high-energy neutron source (INS) (neutron energies $\sim$ 14~MeV). The neutrons
are emitted by fusion reactions during compression of a D-T gas by cumulative
detonation waves. The system of equations for the INS includes: the system of
gas dynamic equations that takes into account the radiation energy transfer;
the equations for the radiation flux; the equation of the shock adiabat
(Hugoniot adiabat) for a compressed gas; the equation for the neutron yield.

We performed the simulation of INS dynamics for spherical and cylindrical
geometries and estimated the maximum temperatures of D-T plasma, its density
and neutron yield per pulse.

The obtained estimates and simulation results show that on the basis of this
approach, the thermonuclear fusion temperatures are reached, and a fusion
reaction proceeds. The yield of this reaction determines the neutron yield.

It should be noted that the feasibility of INS is also confirmed by the
published experimental results~\cite{1,2,3,4,5}. The deviation of the
experimental results from the idealized calculated neutron yields
in~\cite{1,2,3,4,5} may be related to the following reasons.

The problem of asymmetry of a convergent spherical shock wave: a complex
system of fuses was used to initiate the blast wave from the explosive surface.
The shock wave asymmetry could be caused by the discrete zones of the fuses,
and by the errors in the electronic fuse syncronizer.

The small neutron yield can also be associated with the small amount of D-T
mixture (only 0.1~g) and insufficient amount of explosive (there is no data on
the amount of explosive in the experimental device in~\cite{1,2,3,4,5}). The
neutron yield is determined by the number of fusion reactions. As shown
in~\cite{1}, this number is determined by the ratio of the fusion reaction
initialization time ${t_{fusion}}$ in compressed D-T mixture and the time of
its inertial compression ${t_{in}}$. After the inertial compression time
${t_{in}}$, starting from the point of maximum compression, the compressed
mixture begins to expand, i.e. scatter in the opposite direction. So for a
significant yield of neutrons, it is necessary to fulfill the condition
${t_{fusion}} \le {t_{in}}$. And the time of inertial compression ${t_{in}}$
depends on the mass of the compressed gas and the explosive in the device. The
mass of explosive in turn depends on the density of the explosive.

As also shown in~\cite{1}, the neutron yield depends on the maximum degree of
compression, since the initialization time of the fusion reaction
${t_{fusion}}$ depends on the compression ratio exponentially, and sharply
decreases with its increase (according to the estimate in~\cite{1}, the  10~\%
increase in the maximum compression decreases ${t_{fusion}}$ by two orders of
magnitude). The maximum degree of compression may be increased using a cascade
of two shock waves.

In order to increase the yield of neutrons and optimize the geometric
parameters and weight of the INS, the R\& D work should focus on the following:
\begin{enumerate}

\item[a)] the compression of the D-T mixture must be realized by means of a
cascade shock wave or an interlaced ("sandwitch") explosive medium;

\item[b)] the synchronization of the shock wave initiation from the surface of
an explosive layer can nowadays be implemented using a laser radiation with a
given wavelength and the explosive sensitive to such
radiation~\cite{17,18,19,20,21,21a,22,23}. This would also solve the problem of
the converging shock wave asymmetry and thus increase its stability;

\item[c)] in order to reduce the radiation losses, it is possible to add some
heavy elements to the composition of the explosives and gas mixture -- as
the passive admixtures. The heavy metal salts, organometallic compounds, as
well as the lead azide, mercury fulminate and other explosives containing heavy
elements. It is also possible to inject the heavy elements into the gas mixture
(like the mercury vapor), or to blow up a small charge of lead azide at the
center at the moment of explosion, or to evaporate a small amount of metal by
explosion~\cite{1};

\item[d)] in order to increase the neutron yield, it is necessary to increase
the amount of reacting deuterium and tritium. For this purpose it is possible
to use the explosives in which hydrogen is chemically replaced by deuterium and
tritium (for example, if one uses the tetranitromethane mixture in nitrobenzene
solution, as in experiment~\cite{1,2,3,4,5}, a heavy nitrobenzene may be used
instead of the usual nitrobenzene). The products of such explosion, which forms
the detonation wave, will contain deuterium and tritium, which will also take
part in the fusion reaction~\cite{1}.
\end{enumerate}

In conclusion, let us note that the method~\cite{6} used to estimate the
temperature of the thermonuclear plasma in the experiment~\cite{2,3,4,5} is
based on the "Brysk Gaussian"~\cite{6}, which determines the neutron energy
distribution and does not take into account the moderation of neutrons in the
medium. So this method may require some further improvement, in particular,
by the recent neutron moderation theory~\cite{24}.

\end{document}